# Is Second Law of Thermodynamics Violated for Electron Transition from Lower-Energy Level to Higher-Energy Level

Dr. R.C. Gupta
Professor & Head, Mechanical Engineering Dept.,
Institute of engineering & Technology (I.E.T.),
Lucknow, India
rcg_iet@hotmail.com

Ruchi Gupta
Graduate Student, Electrical Engg. Dept.,
Stanford University, USA.

Sanjay Gupta
Consultant, Avesta Computers
California, USA

## ABSTRACT

Second law of thermodynamics is applied to a few electronic processes. It is seen that the second law of thermodynamics holds good for all except one mentioned here. The classical approach, based on exact equivalence of emission and absorption spectra, for electron transition from lower-energy level (first orbit) to higher- energy level (second orbit) ($h\nu = E_2 - E_1$) violates the second law of thermodynamics. But since second law which implies irreversibility and is universally true; a new explanation of electron transition from lower to higher energy level is proposed which leads to better understanding of several topics such as Fraunhofer lines, Optical laser. Also, interestingly, it is shown that widely different fields such as 'second law of thermodynamics' and 'special relativity' are in fact closely linked to each other. Also, possible links between 'supersymmetry' and new concept of 'quaternion-mass' are mentioned.

## 1. INTRODUCTION

Second law of thermodynamics implies that 'although work can be fully converted to heat, but heat can not be fully converted to work'. For conversion of heat to work: out of heat supplied $Q_1$ from source, only a part of the heat is converted to work W and the remainder heat $Q_2$ ( > 0 ) is wasted to surrounding (or sink). Mathematically,

$$Q_1 - Q_2 = W \qquad (1)$$

Efficiency of the process is given by,

$$\eta = W/Q_1 = (Q_1 - Q_2)/Q_1 \qquad (2)$$

The second law also leads to conclusion of 'irreversibility'. It may be noted that although the second law of thermodynamics is generally used for study of thermodynamic processes, the law seems to be universally valid and no exception has been found as yet except the one reported in this paper.

In the present paper; it is shown that when the validity of second law of thermodynamics is tested to a few electronic processes, it works very well except the one mentioned here. It is noted that the current approach of electron transition from first orbit to second orbit ($h\nu = E_2 - E_1$) based on the assumption that emission spectra & absorption spectra are 'exactly' same, violates second law of thermodynamics. So a new possible explanation of electron transition from lower to higher orbit is suggested in this report in accordance with the second law. Interestingly, it is found that two differently looking subjects 'second law of thermodynamics' and 'special relativity' are inter-linked. One can be derived from other, and fall of one means fall of the other. In view of vast literature available in libraries and on internet on thermodynamics, relativity and electronic processes, the author prefers to quote only a few general books [1-5].

## 2. HEAT AND WORK

In thermodynamic processes 'heat' and 'work' are generally obvious, but there is some misconception too. The so-called heat of a hot body, as per second law of thermodynamics, is in fact not 'heat' but 'work' due to vibration/motion of atom/molecules. In electronic processes where usually 'energy' transfer/transition takes place, recognition of heat & work is even more difficult. What is energy? Is energy 'heat' or 'work'? A little consideration, however, will reveal that all energies such as potential energy, kinetic energy, electrostatic energy, chemical energy, nuclear energy, mass energy $mc^2$ etc. in a way are 'work', except the 'radiation energy $h\nu$' which is 'heat'. In fact 'heat' is the energy carried by mass-less particle such as photon whereas energy carried by massive particle is 'work'. In other words; boson (photon) carries the 'heat' as radiation waves, whereas fermion (electron or fermion-groups as atoms/molecules) carries 'work' as kinetic & potential energy of particle. In the following electronic processes these concepts of 'heat ' and 'work' is used to test the validity of second law of thermodynamics.

## 3. SECOND LAW OF THERMODYNAMICS APPLIED TO ELECTRONIC PROCESSES

Electronic processes under consideration for testing the validity of second law of thermodynamics are grouped in two categories as mentioned below and are discussed in sections 3.1 and 3.2 respectively:

**1.** Electronic processes where 'work is converted to heat' ; second law of thermodynamics allows full (100%) conversion of work into heat.

**2.** Electronic processes where 'heat is converted to work' ; as per second law of thermodynamics only a partial (<100%) conversion of heat into work is possible.

### 3.1 Work to Heat

#### 3.1.1 Heating of electric wire by passing current

The energy supplied V.i.t delivered towards kinetic energy of electrons finally at steady-state situation converts fully to radiation-heat.

#### 3.1.2 X-ray

The kinetic energy of electron ($\frac{1}{2} mv^2 = e.V$) may be fully converted to X-ray radiation (hν) of min. wavelength λ (=h.c/(e.V)) .

#### 3.1.3 Annihilation

A particle (say, electron) and an antiparticle (say, positron) can annihilate each other giving two γ-ray photons i.e., $e^- + e^+ = \gamma + \gamma$ . The total mass energy 1.02 MeV plus kinetic energy of the particles converts 100% to the energy of photons.

#### 3.1.4 Electron transition from higher orbit to lower orbit

The energy of an electron in second orbit is $E_2$ and in first orbit is $E_1$. When the electron jumps down from second orbit to first orbit, the difference of energy (work) $E_2 - E_1$ is fully converted to radiation (heat) energy $h\nu_{12}$.

Work to Heat     100% conversion (work to heat) possible i.e.,

$$E_2 - E_1 = h.\nu_{12} \qquad (3)$$

### 3.2 Heat to Work

#### 3.2.1 Photoelectric effect

The photo-electric equation $h\nu - h\nu_o = \frac{1}{2} mv^2$ is indeed in accordance with the second law of thermodynamic equation $Q_1 - Q_2 = W$ , which means that the work function ($h\nu_o$) corresponding to $Q_2$ can never be zero. The efficiency of the process is therefore $\eta = (\nu - \nu_o)/\nu < 1$ .

#### 3.2.2 Compton effect

The Compton effect equation for energy conservation (first law of thermodynamics) $h\nu - h\nu' = \frac{1}{2} m_o v^2$ or more precisely ,

$$h\nu - h\nu' = m_o c^2/(1 - v^2/c^2)^{1/2} - m_o c^2 \quad (4)$$

is also in accordance with second law of thermodynamics equation $Q_1 - Q_2 = W$, which means that $\nu'$ can never be zero. The efficiency of the process is, similar to that for photo-electric effect, therefore

$$\eta = (\nu - \nu')/\nu < 1 \quad (5)$$

### 3.2.3 Pair-production

If a γ-ray of energy (heat) 1.02 MeV could produce in empty space the electron positron pair $2mc^2$ (work) then it would be 100% conversion of heat to work against the second law of thermodynamics; but it is known that this does not happen. In fact some other object such as nucleus is involved in the pair-production process to carry away part of initial photon's energy, thus only less than 100% of photon's energy (heat) is utilized in producing the electron positron pair $2mc^2$ (work). The author opines/predicts that : if, however, pair-production would ever to happen in empty space, then it would be like $\gamma = e^- + e^+ + \gamma_o$ giving out a low energy photon ($\gamma_o$) such that γ should be more than 1.02 MeV according to second law of thermodynamics.

### 3.2.4 Electron transition from lower orbit to higher orbit

The classical approach suggests that if the radiation (heat) energy $h\nu_{12}$ is given to an electron in the first orbit then it should jump up to the second orbit, utilizing it for energy (work) increase $E_2 - E_1$ implying that

$$h.\nu_{12} = E_2 - E_1 \quad (6)$$

i.e., Heat to Work    100% conversion (heat to work) possible

But this *NOT POSSIBLE* as 100% conversion of Heat to Work is strictly prohibited by second law of thermodynamics

## 4. THERMODYNAMICS AND RELATIVITY LINKED TOGETHER

Consider the Compton effect again. The outgoing photon ought to exist, i.e., $\nu'$ can never be zero (which is in accordance with second law of thermodynamics) is analyzed further considering the two possibilities :

(i) If the incident photon strikes the electron in x-direction and if after collision electron deviates from x-direction, photon ought to come out to balance the momentum of electron in y-direction.

(ii) If the incident photon strikes the electron in x-direction and if after collision electron too moves in x-direction, there is no momentum in y-direction ; there are two possibilities
 (a) Photon after impact may be absent ($v' = 0$) or
 (b) Photon after impact may come out ($v' > 0$) in x-direction forward or backward.

But the second law of thermodynamics ($Q_2 > 0$) to hold good the possibility (a) is ruled out and possibility (b) is to occur, $v' > 0$, thus the energy equation (eq.4) for Compton effect written earlier can be re-written using $v' > 0$ as follows,

$$h\nu > m_o c^2/(1 - v^2/c^2)^{1/2} - m_o c^2 \qquad (7)$$

and the conservation of momentum in x-direction yields,

$$h\nu = m_o/(1 - v^2/c^2)^{1/2} \cdot v \cdot c \qquad (8)$$

Putting value of $h\nu$ from Eq.(8) into Eq.(7) and after a few steps of simplification, the following interesting result is found to emerge out as

$$v < c \qquad (9)$$

which is well in accordance with the theory of Relativity.

This means that the result of second law of thermodynamics ($v'$ is never zero, or heat to work conversion $\eta = W/Q_1 < 1$) is compatible with the essence of special relativity ($v < c$ i.e.; no matter how energetic may be the incident photon, velocity of the electron can not exceed velocity of light c, implying $\beta = v/c < 1$). It is amazing that how two quite different fields of study - 'Thermodynamics ($\eta < 1$)' and 'Relativity ($\beta < 1$)' are inter-supportive and inter-linked to each other and appear to be two faces of the same coin.

## 5. IS SECOND LAW OF THERMODYNAMICS REALLY VIOLATED FOR ELECTRON TRANSITION FROM LOWER ORBIT TO HIGHER ORBIT

Although the processes explained in section 3.1.1, 3.1.2, 3.1.3, 3.2.1, 3.2.2 support the second law of thermodynamics; but as explained in section 3.2.3 the second law of thermodynamics is 'violated' if the existing concept (Eq.6) is assumed to be true or in other words if absorption-spectra frequency is considered to be 'exactly' equal to the emission-spectra frequency for corresponding transition. But second law of thermodynamics which implies irreversibility is universally true. Moreover the second law of thermodynamics is closely linked to well established special-relativity. Fall of second law of thermodynamics means fall of relativity. The author opines that- to come out of this crisis there is a need for new explanation of electron transition from lower energy-level to higher energy-level which must be in accordance with the second law of thermodynamics. The new explanation is proposed as following.

## 6. NEW EXPLANATION FOR ELECTRON TRANSITION FROM LOWER TO HIGHER ENERGY-LEVEL (ORBIT)

Consider an electron in the first orbit with energy $E_1$. The energy levels of second and third orbits are $E_2$ and $E_3$ respectively. Although $\nu_{13}$ is the frequency of radiation that would be emitted if the electron comes from third orbit to first orbit; but if the radiation 'exactly' equal to $\nu h_{13}$ is given to the electron in the first orbit, it would not go to third orbit because this too would 'violate' the second law of thermodynamics as mentioned in section 3.2.3 for electron transition from first to second orbit. However, the electron of the first-orbit after receiving the radiation $h\nu_{13}$ can easily go to second orbit absorbing a part ($h\nu_{12}$) of the supplied energy ($h\nu_{13}$) and rejecting the remainder ($h\nu_{13}-h\nu_{12}$), in accordance with the second law of thermodynamics,

$$Q_1 = W + Q_2 \quad \text{or},$$

$$h\nu_{13} = E_2 - E_1 + h(\nu_{13} - \nu_{12})$$

Efficiency of the process in this case would be, from Eq.2 is given as:

$$\eta = [\{h\nu_{13} - h(\nu_{13} - \nu_{12})\}/(h\nu_{13})] = \nu_{12}/\nu_{13} \tag{10}$$

But consider what happens if radiation $h\nu_{12}$ falls upon electron of the first orbit. It can not go to the second orbit otherwise second law of thermodynamics would be violated as mentioned in section 3.2.3. It can, however, go to a 'meta-stable energy level M' between first & second orbits, absorbing a part ($h\nu_{1M}$) of the supplied energy ($h\nu_{12}$) and rejecting the remainder ($h\nu_{12} - h\nu_{1M}$). Again; since $Q_1 = W + Q_2$ and $\eta = W/Q_1$,

$$\eta = [\{h\nu_{12} - h(\nu_{12} - \nu_{1M})\}/h\nu_{12})] = \nu_{1M}/\nu_{12} \tag{11}$$

## 7. CONSEQUENCES OF THE NEW EXPLANATION OF ELECTRON TRANSITION FROM LOWER TO HIGHER ENERGY LEVEL

The new explanation, which is in accordance with the second law of thermodynamics has far reaching consequences for better understanding of Fraunhofer lines, Laser, Fluorescence & phosphorescence briefly described as follows.

### 7. 1 Fraunhofer Line

Fraunhofer lines are the dark lines in the continuous solar spectra and are considered to be present due to absorption of certain radiations into the few elements present in the atmosphere. Since absorption and emission are now shown to be 'not exactly same but different' in this paper, in-principle a re-identification of the elements present in the atmosphere corresponding to Fraunhofer lines may be needed. Although the emission-spectra and absorption-spectra are be different, but possibly the difference may be so little to be noticed.

**7. 2 Laser**

Consider ruby-crystal optical laser. In optical laser conventionally it is considered that pumping (for population inversion) is done as follows: electrons from the first orbit goes to second and third orbits after absorbing radiations $h\nu_{12}$(yellow) and $h\nu_{13}$(green) respectively, and after losing some energy to lattice come down to meta-stable state M which can then liberate $h\nu_{1M}$(red) if the electrons from M-state fall back to the first orbit. But according to the present explanation (as per section 6), the electrons in the first orbit receiving $h\nu_{12}$(yellow) instead of going to second orbit directly goes to meta-stable state; and receiving $h\nu_{13}$ do not go to the third orbit but after some rejection ($h\nu_{13}-h\nu_{12}$ considered to be used up for lattice) go up to second orbit from where the electron eventually comes to the first orbit radiating $h\nu_{12}$. This ($h\nu_{12}$) in turn sends another electron to meta-stable state M which can finally liberate $h\nu_{1M}$(red) if the electron falls back to the first orbit.

**7. 3 Fluorescence and Phosphorescence**

On similar lines as explained in section 7.2 for optical laser, it is suggested that for fluorescence and phosphorescence too, electrons in the meta-stable state do not come from higher orbit but reach directly from first orbit.

**8. POSSIBLE THERMODYNAMIC LINKS OF SUPERSYMMETRY TO QUATERNION MASS**

**8.1 Supersymmetry**

Supersymmetry means that there is transformation which relates the particles of integral spin such as photon (boson) to the particles of half-integral spin such as electron (fermion). Bosons are the 'mediators' of the fundamental forces while fermions make up the 'matter'. The supersymmetry solves the 'hierarchy problem' for grand-unification. Also, for unification of forces, with supersymmetry the promising string-theory becomes the better & famous superstring theory [6]. But as yet, no supersymmetric-partner particles have been found. An ambitious attempt is made, as follows, to answer : 'why superpartners exist in principle but not in reality'.

**8.2 Quaternion Mass**

The author in another interesting paper [7] has proposed a novel concept of mass, i.e., mass of a moving body is neither scalar nor vector but rather 'mass is a quaternion

(scalar + vector) quantity' as $M = m_g + m_p$ where $m_g$ is the grain(scalar)-mass governing the particle-behavior & $m_p$ is the photonic(vector)-mass exhibiting the wave-properties and the total magnitude-of-mass [M] is the Pythagorean sum as $[M] = m_g^2 + m_p^2$ . The proposed concept of 'Quaternion Mass' gives a mathematical blend to the de-Broglie hypothesis and explains well the wave-particle duality. Also, in a way, it unifies the quantum-mechanics and the special-relativity.

## 8.3 Links between Quaternion-Mass and Supersymmetry

Referring to the 'quaternion mass concept', it may be noted that the grain-mass $m_g$ signifies the material content due to atoms/molecules (group of fermions) whereas the photonic-mass $m_p$ is due to associated photon (boson). So, in a way, the quaternion mass $M = m_g + m_p$ is 'marriage' of fermiom plus photon. Thus the supersymmetry seems to be inherently embedded in the quaternion mass concept. It is as if, fermion & boson are coupled to each other and that fermion behaves as particle & boson shows the wave nature. Wave-particle duality seems, thus in a way, due to supersymmetry embedded in quaternion mass.

It is mentioned in section 2 'Heat & Work' of this paper, that thermodynamically 'heat' is carried by 'boson' whereas 'work' is carried by 'fermion'. Also it is mentioned in section 3 that as per second-law of thermodynamics, although all 'work' can be coverted to 'heat' but all 'heat' can-not be converted to 'work'. This (second-law) thus implies that although all fermions(carrying 'work') may have its superpartner bosons(carrying 'heat') but all bosons(carrying 'heat') may-not have its superpartner fermion(carrying 'work'). Thus, although the supersymmetry is embedded into the concept of quaternion mass but the second-law of thermodynamics limits the supersymmetry i.e., all fermions may have its superpartner bosons but all bosons will-not have its superpartner fermions. Logically it can be argued that, therefore, the existence of supersymmetric-partners (of fermion & boson , both) and thus the supersymmetry itself is doubtful.

It seems that supersymmetric partners (& thus the supersymmetry) can exists only in 'married' state as quaternion sum as $M = m_g + m_p$ and that their separate existence is forbidden by the second-law of thermodynamics.

## 9. DISCUSSIONS

Heat is considered to be an averaged quantity, so question arises whether applying second law of thermodynamics to single photon inter-action such as Compton-effect etc. is reasonable or not ? The author opines that it is indeed reasonable because: (i) average of a single data average is the data itself, a photon of energy hν has its average energy too as hν , (ii) who says or what prohibits that the second law of thermodynamics can not be applied to single photon interaction system, (iii) when the second law is applied to single photon interaction such as Compton-effect, it not only holds good very well but also shows new light(understanding) i.e., it leads to conclusion that not only the

Compton-effect is in accordance with the second law of thermodynamics but also that the law of Thermodynamics is in accordance with the theory of Relativity, (iv) Theory of relativity is applicable well to single particle so is the second law of thermodynamics, as both are shown to be linked, (v) all avenues for Truth must be kept open; when the second law is applied to the electronic processes or elsewhere described in the paper, its success is striking .

## 10. CONCLUSIONS

Although many electronic processes considered are in support to the second law of thermodynamics but conventional approach ($h\nu_{12} = E_2 - E_1$) of electron transition from lower to higher orbit violates the second law of thermodynamics. Thus a new explanation of electron transition from lower to higher orbit, in accordance with the second law is suggested, which leads to possibly fine re-defining of the Fraunhofer lines and better understanding of optical laser, fluorescence & phosphorescence. Interestingly, special-relativity and second-law of thermodynamics are found to be closely linked. Also, it seems that the supersymmetry is embedded in coupled-state into the quaternion-mass and that the second-law of thermodynamics prohibits the separate existence of the superpartners.

ACKNOWLEDGEMENT
The author is thankful to Dr.M.S.Kalara, Professor, I.I.T.Kanpur and Dr. V. Johri. Prof. Emeritus, Lucknow University for their advice & discussions. The author is also thankful to Veena, Shefali, Dr.Chhavi Gupta & Dr.Sanjiv Sahoo for their assistance.